\documentclass[11pt,twoside]{article}

\usepackage{pstricks,pst-coil,multicol}
\psset{linewidth=0.048cm}\psset{unit=0.8cm}

\newcommand{\be}{\begin{equation}}
\newcommand{\ee}{\end{equation}}
\newcommand{\bea}{\begin{eqnarray}}
\newcommand{\eea}{\end{eqnarray}}
\newcommand{\beas}{\begin{eqnarray*}}
\newcommand{\eeas}{\end{eqnarray*}}
\newcommand{\bdm}{\begin{displaymath}}
\newcommand{\edm}{\end{displaymath}}
\newcommand{\ba}{\begin{array}}
\newcommand{\ea}{\end{array}}
\newcommand{\bi}{\begin{itemize}}
\newcommand{\ei}{\end{itemize}}
\newcommand{\ben}{\begin{enumerate}}
\newcommand{\een}{\end{enumerate}}
\newcommand{\bc}{\begin{center}}
\newcommand{\ec}{\end{center}}
\newcommand{\bfl}{\begin{flushleft}}
\newcommand{\efl}{\end{flushleft}}
\newcommand{\bfr}{\begin{flushright}}
\newcommand{\efr}{\end{flushright}}
\newcommand{\bd}{\begin{description}}
\newcommand{\ed}{\end{description}}
\newcommand{\bq}{\begin{quote}}
\newcommand{\eq}{\end{quote}}
\newcommand{\bfg}{\begin{figure}}
\newcommand{\efg}{\end{figure}}
\newcommand{\bt}{\begin{table}}
\newcommand{\et}{\end{table}}
\newcommand{\btb}{\begin{tabular}}
\newcommand{\etb}{\end{tabular}}
\newcommand{\btg}{\begin{tabbing}}
\newcommand{\etg}{\end{tabbing}}
\newcommand{\bp}{\begin{picture}}
\newcommand{\ep}{\end{picture}}

\newcommand{\NM}{\normalsize}
\newcommand{\SI}{\small}
\newcommand{\SII}{\footnotesize}

\newcommand{\bm}[1]{\mbox{\boldmath $#1$}}
\newcommand{\itg}{\int \limits}

\newcommand{\eps}{\varepsilon}

\renewcommand{\Im}{\mbox{Im} \,}

\newcommand{\titlestyle}{\em}
\newcommand{\volstyle}{\bf}

\newcommand{\Li}{\mbox{Li}_{2}}
\newcommand{\Tri}{\mbox{Li}_{3}}
\newcommand{\Spe}{\mbox{S}_{12}}

\newcommand{\prepname}{(in preparation)}

\newcommand{\mcl}{\cal}
\newcommand{\nop}{\rule{0cm}{0cm}}

\begin{document}

\title{
\bfr \NM MZ-TH/97-23 \\[1cm] \efr 
\bf {\it oneloop 2.0} -- A Program Package \\[1cm]
calculating One-Loop Integrals \\[2cm]}
\author{L. Br\"ucher${}^{1}$,
J. Franzkowski${}^{2}$, D. Kreimer${}^{3}$ \\[0.5cm]
{\NM Institut f\"ur Physik} \\
{\NM Johannes Gutenberg-Universit\"at} \\
{\NM Staudingerweg 7} \\
{\NM D-55099 Mainz} \\
{\NM Germany} \\[1.5cm]} 
\date{May 1997 \\[1cm]} 

\maketitle

\begin{abstract}

We present an improved version of our program package {\it oneloop} which --
written as a package for MAPLE \cite{Ma1,Ma2} -- solves one-loop Feynman
integrals. The package is calculating one-, two- and three-point functions both
algebraically and numerically to any tensor rank. In addition to the original
version {\it oneloop 2.0} also calculates infrared divergent integrals. Higher
powers of propagator terms and the ${\mcl O}(\eps)$ parts relevant for two-loop
calculations are now supported.

\end{abstract}

\footnotetext[1]{e-mail: \tt bruecher@dipmza.physik.uni-mainz.de}
\footnotetext[2]{e-mail: \tt franzkowski@dipmza.physik.uni-mainz.de}
\footnotetext[3]{e-mail: \tt kreimer@dipmza.physik.uni-mainz.de}

\thispagestyle{empty}

\newpage

\setcounter{page}{1}

\section*{NEW VERSION SUMMARY}

{\small {\it Title of new version: } oneloop.ma, version 2.0 \\[0.4cm]
{\it Reference to original program: } oneloop.ma \\[0.4cm]
{\it Reference in CPC: } Comput. Phys. Commun. 85 (1995) 153 \\[0.4cm]
{\it Does the new version supersede the old program?} Yes \\[0.4cm]
{\it Licensing provisions:} None \\[0.4cm]
{\it Computers: } Any platform which has MAPLE V, Release 3 \\[0.4cm]
{\it No. of processors used: } 1 \\[0.4cm]
{\it Operating system under which the new version has been tested: } (i)
Unix (Linux 2.0) (ii) VMS 6.2 (iii) MS-DOS 6.0 \\[0.4cm]
{\it Programming language: } MAPLE V, Release 3 \\[0.4cm]
{\it Subprograms used: } cfcn.ma, r.ma, simple.ma, pv.ma, mess.ma \\[0.4cm]
{\it Memory required to execute with typical data: } Less than 8 MB RAM \\[0.4cm]
{\it Typical running time: } The algebraic or numeric evaluation of a three-point 
function up to the rank-3-tensor on a DEC Alpha 8400 needs less than 2s \\[0.4cm]
{\it No. of lines in distributed program including subprograms: } 4019 \\[0.4cm]
{\it No. of bytes in distributed program including test data: } 168 189 \\[0.4cm]
{\it Distribution format: } binary (zipped ASCII files) \\[0.4cm]
{\it Keywords: } Electroweak theory, Feynman diagrams, one-loop corrections, 
renormalization \\[0.4cm]
{\it Nature of physical problem: } The theoretical determination of cross 
sections in particle processes requires the calculation of radiative 
corrections. The most important contribution comes from the level of 
one-loop Feynman diagrams which arise from the model under consideration, 
usually the standard model of elementary particles \\[0.4cm]
{\it Method of solution: } This package is designed to evaluate 
automatically one-loop integrals. It is making use of the properties of 
${\cal R}$ functions, a class of special functions which simplifies the 
evaluation of Feynman integrals}

\section*{LONG WRITE-UP}

\section{Introduction}

In particle physics the calculation of one- and higher loop corrections to
particle processes is mandatory to keep track with the increasing accuracy of
particle colliders. For this reason the necessary calculations were automated by
virtue of computer programs during recent years. In this context the program
{\it oneloop} was developed at the one-loop level \cite{Fr4}. It provides the
necessary integrals which are needed for Feynman diagrams with at most three
external legs.

Several improvements of the original version now merged in a new version 2.0. It
contains the following new features:
\bi
\item Arbitrary powers of the propagator terms in the denominator are allowed
  (cf. sect.~\ref{power}). The original version was restricted to propagators of
  power one.
\item Infrared divergences can be calculated in dimensional regularization. They
  will appear as poles in the dimensional regulator $\eps=(4-D)/2$ like in the
  ultraviolet case (cf. sect. \ref{infra}).
\item The contributions of ${\mcl O}(\eps)$ can be calculated as well. Although
  these terms are not contributing in pure one-loop calculations they become
  relevant in two-loop applications (cf. sect. \ref{eps}).
\item Several kinematical points in which the original version ran into
  numerical instability or terminated in a division by zero are now solved in a
  different, numerically stable way (cf. sect. \ref{num}).
\item Additional functions {\tt OneLoopTens}{\it n}{\tt Pt} accept squared
  momenta as input and return full tensors (cf. sect. \ref{tens}).
\item The usage of a library of one-loop integrals is generalized (cf. sect.
  \ref{lib}).
\item Some other minor changes have to be reported (cf. sect. \ref{misc}):
  
  In the case of the three-point function the convention of the output changed
  slightly. It has now the same structure as for the other functions.
  
  Some new functions which abbreviate the output are introduced in the case
  where the functions are calculated analytically.
\ei
This note describes all changes in more detail.

\section{Arbitrary powers of propagators}\label{power}

In addition to the original functions
\bi
\item {\tt OneLoop1Pt}$(p,m)$
\item {\tt OneLoop2Pt}$(p_0,p_1,q,m_1,m_2)$
\item {\tt OneLoop3Pt}$(p_0,p_1,p_2,q_1,q_{20},q_{21},m_1,m_2,m_3)$
\ei
-- which need the tensor degree, the external momenta $q_n$ and the masses
$m_n$ as input -- there exist now the following generalizations
\bi
\item {\tt OneLoop1Pt}$(p,m,t)$
\item {\tt OneLoop2Pt}$(p_0,p_1,q,m_1,m_2,t_1,t_2)$
\item {\tt OneLoop3Pt}$(p_0,p_1,p_2,q_1,q_{20},q_{21},m_1,m_2,m_3,t_1,t_2,t_3)$.
\ei
These functions additionally expect the powers $t_n$ of the propagator terms in
the denominator. In detail they directly correspond to the following integrals:
\bi
\item One-point function: \\
      \parbox{10.6cm}{\bc\SII
      \parbox{10cm}{\bp(10,4)
      \pscircle(5,2.5){1.5}
      \psline{->}(6.47,2.6)(6.47,2.4)
      \pscircle[linewidth=0.1](3.53,2.5){0.1}
      \psline(1,2.5)(3.5,2.5)
      \put(6.7,2.5){$\bm{l}$}
      \put(5.9,2.5){$m$}
      \ep} 
      \ec} \vspace{-0.5cm} \nop
      \bea
      & & \mbox{\tt OneLoop1Pt}(p,m,t) = A^{(p)(t)}(m) = \int \! d^{D}l \, 
      \frac{(l^2)^{\frac{p}{2}}}{[l^{2} - m^{2} + i \varrho]^t}
      \eea \\[0.5cm] \nop
\item Two-point function: \\
      \parbox{10.6cm}{\bc\SII
      \parbox{10cm}{\bp(10,4)
      \pscircle(5,2.5){1.5}
      \psline{->}(5.1,3.97)(4.9,3.97)
      \psline{->}(4.9,1.03)(5.1,1.03)
      \pscircle[linewidth=0.1](3.53,2.5){0.1}
      \pscircle[linewidth=0.1](6.47,2.5){0.1}
      \psline{->}(1,2.5)(2.45,2.5)
      \psline(2.25,2.5)(3.5,2.5)
      \psline{->}(6.5,2.5)(7.95,2.5)
      \psline(7.75,2.5)(9,2.5)
      \put(4.85,1.3){$m_{1}$}
      \put(4.85,3.6){$m_{2}$}
      \put(4.6,0.53){$\bm{l+q_{1}}$}
      \put(5,4.2){$\bm{l}$}
      \put(2.05,2.77){$\bm{q_{1}}$}
      \put(7.55,2.77){$\bm{q_{1}}$}
      \ep}
      \ec} \vspace{-0.5cm} \nop
      \bea
      \lefteqn{\mbox{\tt OneLoop2Pt}(p_0,p_1,q,m_1,m_2,t_1,t_2) = 
      B^{(p_{0} p_{1})(t_{1} t_{2})}(q, m_{1}, m_{2})} \nonumber \\[0.2cm]
      & = & \int \! d^{D}l \, \frac{(l_{\|})^{p_{0}} \, (l_{\bot})^{p_{1}}}
      {[(l + q)^{2} - m_{1}^{2} + i \varrho]^{t_1} \, [l^{2} - m_{2}^{2} + 
      i \varrho]^{t_2}} \hspace*{2.9cm} \nop
      \eea
      Abbreviations:
      \be
      \label{split1} 
      l_{\|} = \frac{l\cdot q}{\sqrt{q^2}} \,; \qquad
      l_{\bot} = \sqrt{l_{\|}^2 - l^2}
      \ee
\item Three-point function: \\
      \parbox{10.6cm}{\bc\SII
      \parbox{10cm}{\bp(10,4)
      \pscircle[linewidth=0.1](3.53,2.5){0.1}
      \pscircle[linewidth=0.1](6.5,1){0.1}
      \pscircle[linewidth=0.1](6.5,4){0.1}
      \psline{->}(1,2.5)(2.45,2.5)
      \psline(2.25,2.5)(3.5,2.5)
      \psline{->}(6.5,4)(7.95,4)
      \psline(9,4)(7.55,4)
      \psline(7.75,1)(6.5,1)
      \psline{->}(9,1)(7.55,1)
      \psline{->}(3.5,2.5)(5.2,1.65)
      \psline(5.1,1.7)(6.5,1)
      \psline{->}(6.5,4)(4.8,3.15)
      \psline(5.1,3.3)(3.5,2.5)
      \psline{->}(6.5,1)(6.5,2.7)
      \psline(6.5,2.4)(6.5,4)
      \put(4.1,1.35){$\bm{l+q_{1}}$}
      \put(6.715,2.35){$\bm{l+q_{2}}$}
      \put(4.8,3.35){$\bm{l}$}
      \put(4.85,2){$m_{1}$}
      \put(5.8,2.35){$m_{2}$}
      \put(4.85,2.9){$m_{3}$}
      \put(2.15,2.77){$\bm{q_{1}}$}
      \put(7.65,3.6){$\bm{q_{2}}$}
      \put(7.1,1.2){$\bm{q_{2}-q_{1}}$}
      \ep}
      \ec} \vspace{-0.5cm} \nop
      \bea
      \lefteqn{\mbox{\tt OneLoop3Pt}(p_0,p_1,p_2,q_{1},q_{20},q_{21},m_1,
      m_2,m_3,t_1,t_2,t_3)} \nonumber \\[0.2cm]
      & = & C^{(p_{0} p_{1} p_{2})(t_1,t_2,t_3)}(q_{1}, q_{2}, m_{1}, m_{2},
      m_{3}) \\[0.2cm]
      & = & \int \! d^{D}l \, \frac{(l_{0\|})^{p_{0}} \, (l_{1\|})^{p_{1}} \, 
      (l_{\bot})^{p_{2}}}{[(l + q_{1})^{2} - m_{1}^{2} + i \varrho]^{t_1} \, 
      [(l + q_{2})^{2} - m_{2}^{2} + i \varrho]^{t_2} \, [l^{2} - m_{3}^{2} + i 
      \varrho]^{t_3}} \nonumber
      \eea
      Abbreviations:
      \bea
      \label{split2}
      & & l_{0\|} = \frac{l\cdot q_1}{\sqrt{q_1^2}} \,;
      \qquad l_{1\|} = - \frac{l\cdot {q'}_2}{\sqrt{{q'}_2^2}} \,;
      \qquad {q'}_2 = q_2 - \frac{q_1\cdot q_2}{q_1^2}\,q_1 \\[0.2cm]
      & & l_{\bot} = \sqrt{l_{0\|}^2 - l_{1\|}^2 - l^2} 
      \,; \qquad q_{20} = \frac{q_1\cdot q_2}{\sqrt{q_1^2}} \,; \qquad
      q_{21} = \sqrt{q_{20}^2 - q_{2}^2} \nonumber
      \eea
\ei
Our notation for the {\tt OneLoop{\rm\it n}Pt} functions distinguishes between
parallel and orthogonal space which is reflected by the definitions of
(\ref{split1}) and (\ref{split2}) for the momentum components \cite{Col,Fr13}.
Please keep in mind the fact that in this notation the indices $p_0, p_1,
\ldots$ represent the powers of the different components of the loop momentum
$l$ which should not be mixed with Lorentz indices. In our notation
$l_{\|},l_{0\|},l_{1\|}$ describe the components of $l$ which are parallel to
the external momenta $q,q_{1},q_{2}$, whereas $l_{\bot}$ represents the
orthogonal complement.

If the powers $t_n$ of the propagator terms in the denominator are omitted the
program assumes the default value 1 for each $t_n$.

\section{Two-loop relevant parts of one-loop diagrams}\label{eps}

The results of the {\tt OneLoop{\rm\it n}Pt} functions are Laurent expansions in
terms of the ultraviolet regulator $\eps$. Therefore the output of the {\tt
OneLoop{\rm\it n}Pt} procedures consists of a list where the significant
coefficients (${\mcl O}(\eps^{-1})$, ${\mcl O}(\eps^0)$) of this expansion are
given.

In two-loop calculations one-loop diagrams get multiplied with divergent
$Z$-factors. Two-loop integrals can factorize into a product of one-loop
integrals. This is the reason why one-loop contributions of ${\mcl O}(\eps^1)$
are also of interest \cite{Nie,Fr9}. For that purpose this coefficient is
calculated if the qualifier {\tt more} is used in the function call:
\\ \nop
\bc
\btb{|l|l|}
\hline
input & output \\
\hline
{\tt OneLoop1Pt}$(p,m,t)$ & 
$[\eps^{-1}\mbox{-{\em term}},\eps^0\mbox{-{\em term}}]$ \\
\hline
{\tt OneLoop1Pt}$(p,m,t,\mbox{\tt more})$ & 
$[\eps^{-1}\mbox{-{\em term}},\eps^0\mbox{-{\em term}},
\eps^1\mbox{-{\em term}}]$ \\
\hline
{\tt OneLoop2Pt}$(p_0,p_1,q,m_1,m_2,t_1,t_2)$ & $[\eps^{-1}\mbox{-{\em term}},
\eps^0\mbox{-{\em term}}]$ \\
\hline
{\tt OneLoop2Pt}$(p_0,p_1,q,m_1,m_2,t_1,t_2,\mbox{\tt more})$ & 
$[\eps^{-1}\mbox{-{\em term}},\eps^0\mbox{-{\em term}},
\eps^1\mbox{-{\em term}}]$ \\
\hline
{\tt OneLoop3Pt}$(p_0,p_1,p_2,q_{1},q_{20},q_{21},m_1,m_2,m_3,t_1,t_2,t_3)$ &
$[\eps^{-1}\mbox{-{\em term}},\eps^0\mbox{-{\em term}}]$
\protect\footnote{} \\
\hline
{\tt OneLoop3Pt}$(p_0,p_1,p_2,q_{1},q_{20},q_{21},$ & \\
$\hspace*{4cm} m_1,m_2,m_3,t_1,t_2,t_3,\mbox{\tt more})$ & $[\eps^{-1}
\mbox{-{\em term}},\eps^0\mbox{-{\em term}},\eps^1\mbox{-{\em term}}]$ \\
\hline
\etb
\footnotetext[1]{The output of {\tt OneLoop3Pt} changed slightly compared to the
original version (cf. sect. \ref{misc}).}
\\[1cm]
\ec
In the output ``$\eps^{-1}$-{\em term}'' is meant to represent the divergent
part (the coefficient of $1/\eps$) whereas ``$\eps^{0}$-{\em term}'' is
describing the finite part and ``$\eps^{1}$-{\em term}'' the ${\mcl O}(\eps)$
contribution.

As long as the arguments of the {\tt OneLoop} functions are symbols the output
is given algebraically, whereas numbers inserted for the arguments imply a
numerical result.

\section{Infrared divergences}\label{infra}

Infrared divergences are now regulated as well. This kind of divergence may
occur for instance for the three-point function if the momenta are on-shell and
one mass is set to 0. This effect also appears for a two-point function where a
propagator term with vanishing mass is squared. Since the collinear divergent
case is still excluded the divergence will always occur as a pole term of ${\mcl
O}(\eps^{-1})$. It means that the output involves a non-vanishing
{\em$\eps^{-1}$-term} in the notation of sect. \ref{eps}. There is no explicit
distinction made between UV and IR divergences, there is no infrared dimension
parameter $\eps_{\mathrm{IR}}$. Both kinds of divergences are described by $\eps
= (4 - D)/2$. The result for the collinear case is well-known \cite{Gas}. It is
the only case which has a pole of ${\mcl O}(\eps^{-2})$. We excluded this simple
case, but the user can add it easily himself.

\section{Numerical stability}\label{num}

The accuracy and stability of numerical results may suffer from cancellations 
of large, approximately equal dilogarithms. Since MAPLE supports calculations 
of arbitrary precision, it is possible to increase the number of digits to
improve accuracy. Usually we calculated with 20 or 40 digits.

Nevertheless, at some kinematical points the general procedure breaks down since
divisions by 0 or similar errors are encountered. It turns out that these points
always belong to a kinematical arrangement -- for instance equal or vanishing
masses or momenta -- which is simpler than the general case. For this reason,
even if there were no numerical problems -- in the sense of optimizing the code
-- it is advisable to solve the simpler kinematical situation in a faster than
the general way.

This is done completely automatically for the following kinematical points:
\\[0.4cm]
{\tt OneLoop2Pt:}\\[-0.8cm]\nop
\begin{multicols}{2}
  \SI
  \noindent
  $q=0$ \\
  $m_1=0$ \\
  $m_2=0$ \\
  $q=0$, $m_1=m_2$ \\
  $q^2-m_1^2+m_2^2=0$ \\
  $q^2+m_1^2-m_2^2=0$ \\
  $q+m_1+m_2=0$ \\
  $q-m_1+m_2=0$ \\
  $q+m_1-m_2=0$ \\
  $q-m_1-m_2=0$
\end{multicols}
\noindent
{\tt OneLoop3Pt:}\\[-0.8cm]\nop
\begin{multicols}{2}
  \SI
  \noindent
  $q_1=0$ \\
  $q_{21}=0$ \\
  $m_1=0$ \\
  $m_2=0$ \\
  $m_3=0$ \\
  $q_{20}-q_1=q_{21}$ \\
  $q_{20}-q_1=-q_{21}$ \\
  $q_{20}=q_{21}$ \\
  $q_{20}=-q_{21}$ \\
  $q_1=0$, $m_1=m_3$ \\
  $q_{20}=0$, $q_{21}=0$ \\
  $q_{21}=0$, $q_1=q_{20}$ \\
  $q_{21}=0$, $(q_{20}^2-m_2^2)q_1-(q_1^2-m_1^2)q_{20}$ \\
  $\nop\qquad -m_3^2(q_{20}-q_1)=0$ \\
  $q_{20}=0$, $q_{21}=0$, $m_2=m_3$ \\
  $q_{21}=0$, $q_1=q_{20}$, $m_1=m_2$ \\
  $q_1=0$, $m_1=m_3$, $\sqrt{q_{20}^2-q_{21}^2}+m_1+m_2=0$ \\
  $q_1=0$, $m_1=m_3$, $\sqrt{q_{20}^2-q_{21}^2}-m_1+m_2=0$ \\
  $q_1=0$, $m_1=m_3$, $\sqrt{q_{20}^2-q_{21}^2}+m_1-m_2=0$ \\
  $q_1=0$, $m_1=m_3$, $\sqrt{q_{20}^2-q_{21}^2}-m_1-m_2=0$ \\
  $q_{20}=0$, $q_{21}=0$, $m_2=m_3$, $q_1+m_1+m_3=0$ \\
  $q_{20}=0$, $q_{21}=0$, $m_2=m_3$, $q_1-m_1+m_3=0$ \\
  $q_{20}=0$, $q_{21}=0$, $m_2=m_3$, $q_1+m_1-m_3=0$ \\
  $q_{20}=0$, $q_{21}=0$, $m_2=m_3$, $q_1-m_1-m_3=0$ \\
  $q_{21}=0$, $q_1=q_{20}$, $m_1=m_2$, $q_1+m_1+m_3=0$ \\
  $q_{21}=0$, $q_1=q_{20}$, $m_1=m_2$, $q_1-m_1+m_3=0$ \\
  $q_{21}=0$, $q_1=q_{20}$, $m_1=m_2$, $q_1+m_1-m_3=0$ \\
  $q_{21}=0$, $q_1=q_{20}$, $m_1=m_2$, $q_1-m_1-m_3=0$ \\
  $q_1^2-m_1^2+m_3^2=0$ \\
  $q_1^2+m_1^2-m_3^2=0$ \\
  $q_1^2-m_2^2+m_3^2=0$ \\
  $q_1^2+m_2^2-m_3^2=0$ \\
  $q_1^2+m_1^2-m_2^2=0$ \\
  $q_1^2-m_1^2+m_2^2=0$ \\
  $q_{20}^2-m_2^2+m_3^2=0$ \\
  $q_{20}^2+m_2^2-m_3^2=0$ \\
  $q_{20}^2-q_{21}^2-m_2^2+m_3^2=0$ \\
  $q_{20}^2-q_{21}^2+m_2^2-m_3^2=0$ \\
  $(q_{20}-q_1)^2+m_1^2-m_2^2=0$ \\
  $(q_{20}-q_1)^2-m_1^2+m_2^2=0$ \\
  $q_{20}^2-q_{21}^2-m_1^2+m_2^2=0$ \\
  $q_{20}^2-q_{21}^2+m_1^2-m_2^2=0$ \\
  $q_{20}^2-q_{21}^2=m_2^2$, $q_1^2=m_1^2$, $m_3=0$ \\
  $(q_{20}-q_1)^2-q_{21}^2=m_1^2$, $q_{20}^2-q_{21}^2=m_3^2$, $m_2=0$ \\
  $(q_{20}-q_1)^2-q_{21}^2=m_2^2$, $q_1^2=m_3^2$, $m_1=0$
\end{multicols}
Several of these points correspond to different thresholds of the integrals.
There are only two remaining sources of difficulties:
\bi
\item if an integral is first solved in a general manner and then -- in the
  result -- special values are substituted which correspond to one of the
  aforementioned kinematical points. Solution: the values have to be assigned
  from the beginning of the calculation.
\item if the kinematical arrangement is not in, but very close to (less than
  roughly $10^{-10}$ compared to the relevant scale) one of the
  above listed kinematical points. Solution: the value of {\tt Digits} has to
  be increased.
\ei

\section{Tensors}\label{tens}

The functions {\tt OneLoopTens{\rm\it n}Pt} expect squared momenta -- $q^2$ for
two-point functions, $q_1^2, q_2^2$ and $(q_2-q_1)^2$ for three-point functions
-- as arguments and return a full tensor as output -- which is completely
equivalent to the notation of (\ref{split1}) and (\ref{split2}) but perhaps the
more familiar representation \cite{Pas}. They replace the similar {\tt PassVelt}
functions of the original version which now are no longer needed.

The following types of arguments -- here only demonstrated for the two-point
case -- are allowed for all {\tt OneLoopTens{\rm\it n}Pt} functions:
\bi
\item {\tt OneLoopTens2Pt}$(i)$: \\
      The rank $i$ tensor decomposition of the two-point
      function is given. The procedure returns a list consisting in
      different coefficients, which are expressed in terms of the {\tt
      OneLoop{\rm\it n}Pt} tensor integrals, and the defining equation for the
      coefficients, for instance in the case $i=2$:
      \beas
      & & \left[C_{21} = - \frac{\mbox{\tt OneLoop2Pt}(0,2)}{3-2\eps} \, , \, 
      C_{20} = - \frac{\mbox{\tt OneLoop2Pt}(0,2)}{q^2 \, (-3+2\eps)}
      \right. \\ 
      & & \hspace*{2cm} \left. + \frac{\mbox{\tt OneLoop2Pt}(2,0)}{q^2} \, , \,
      C_{20} q_{\mu_1} q_{\mu_2} + C_{21} g_{\mu_1\mu_2} \right]      
      \eeas
\item {\tt OneLoopTens2Pt$(i,\mbox{\tt full})$}: \\
      The function inserts the results of the {\tt OneLoop{\rm\it n}Pt} tensor
      integrals explicitly. Here again the case $i=2$:
      \beas
      & & \left[C_{21} = [\eps^{-1}\mbox{-{\em term}},\eps^0\mbox{-{\em term}}]
      \, , \, C_{20} = [\eps^{-1}\mbox{-{\em term}},\eps^0\mbox{-{\em term}}] \,
      , \, C_{20} q_{\mu_1} q_{\mu_2} + C_{21}
      g_{\mu_1\mu_2} \right]      
      \eeas
\item {\tt OneLoopTens2Pt$(i,q^2,m_1,m_2,t_1,t_2)$}: \\
      The function returns the same as {\tt OneLoopTens2Pt$(i,\mbox{\tt
      full})$}, but expressed in the user defined terms for
      $q^2,m_1$ and $m_2$. Of course numerical values are also allowed. $t_1$
      and $t_2$ are optional. 
\item {\tt OneLoopTens2Pt$(i,q^2,m_1,m_2,t_1,t_2,\mbox{\tt more})$}: \\
      The function returns also the ${\mcl O}(\eps)$ contribution. Again, $t_1$
      and $t_2$ are optional:
      \beas
      & & \left[C_{21} = [\eps^{-1}\mbox{-{\em term}},\eps^0\mbox{-{\em term}},
      \eps^1\mbox{-{\em term}}] \, , \, C_{20} = [\eps^{-1}\mbox{-{\em term}},
      \eps^0\mbox{-{\em term}},\eps^1\mbox{-{\em term}}] \, , \right. \\
      & & \hspace*{2cm} \left. C_{20} q_{\mu_1} q_{\mu_2} + C_{21}
      g_{\mu_1\mu_2} \right]      
      \eeas
\ei

\section{Library}\label{lib}

The usage of a library of integrals was restricted to the {\tt PassVelt}
procedures in the original version. It is now applicable for all {\tt OneLoop}
and {\tt OneLoopTens} procedures, so that in any case the procedures speed up.
The package now includes the functions
\bi
\item {\tt OneLoopLib2Pt}($i,j$)
\item {\tt OneLoopLib3Pt}($i,j$)
\ei
which generate a library of all {\tt OneLoop2Pt} and {\tt OneLoop3Pt} functions
respectively up to the tensor rank $i$. All powers of the denominators from 1 to
$j$ are considered. The second parameter $j$ is optional. If it is omitted the
procedures assume the value 1 for $j$, so that no powers of denominators higher
than one are calculated.

If this library shall be written in any other than the current directory one
has to assign the variable {\tt LibPath} which substitutes the variable {\tt
tensorpath} in the original version \\[0.4cm]
\verb+LibPath:=+{$\langle$\it path$\rangle$}\verb+;+ \\[0.4cm]
Now the library will be written to $\langle${\it path}$\rangle$. The procedures
will only look for the library if {\tt LibPath} is assigned. If the procedures
search for an integral which is not contained in the library, an error message
is returned.

In practice it is necessary to store not only the general mass case of each
integral. Several special cases mentioned in sect. \ref{num} are also needed.
The necessary integrals are automatically created by the {\tt OneLoopLib}
routines. Each integral corresponds to one file.

\section{Minor changes}\label{misc}

We changed the convention for the three-point function compared with the
original version: The vanishing $\eps^{-2}\mbox{-{\em term}}$ which was kept for
testing reasons is dropped now. The list of abbreviations which came with the
three-point function also disappeared, because it is no longer needed.

Instead the output now is written in the algebraic case by using several new
abbreviations:
\bea
\mbox{\tt R2ex1}(x,y) & = & \sqrt{1-\frac{x}{y}} \,\, \left[\ln
\left(1-\sqrt{1-\frac{x}{y}}\,\right) - \ln \left(1+\sqrt{1-\frac{x}{y}}\,
\right) + i \pi \right] \\[0.4cm]
& & - \ln(-x) - i \pi \nonumber \\[0.4cm]
\mbox{\tt R2ex2}(x,y) & = & \left(1+\sqrt{1-\frac{x}{y}}\,\right) \, \Li\left(1-
\frac{1-\sqrt{1-\frac{x}{y}}}{1+\sqrt{1-\frac{x}{y}}}\,\right) \nonumber \\
& & +\left(1+\sqrt{1-\frac{x}{y}}\,\right) \, \left[\ln\left(1-\sqrt{1-
\frac{x}{y}}\,\right)\right]^2 \nonumber \\
& & +\left(1-\sqrt{1-\frac{x}{y}}\,\right) \, \Li\left(1-
\frac{1+\sqrt{1-\frac{x}{y}}}{1-\sqrt{1-\frac{x}{y}}}\,\right) \nonumber \\
& & +\left(1-\sqrt{1-\frac{x}{y}}\,\right) \, \left[\ln\left(1+\sqrt{1-
\frac{x}{y}}\,\right)\right]^2 \\[0.4cm]
& & +\frac{1}{2} \, \left(\ln y\right)^2 +2 \, \left(\ln x\right)^2 -2 \, \ln
x \, \ln y \nonumber \\[0.4cm]
& & +\left(\ln y - 2 \, \ln x \right) \, \left[\left(1+\sqrt{1-\frac{x}{y}}\,
\right) \, \ln\left(1-\sqrt{1-\frac{x}{y}}\,\right)\right. \nonumber \\
& & \left. + \left(1-\sqrt{1-\frac{x}{y}}\,
\right) \, \ln\left(1+\sqrt{1-\frac{x}{y}}\,\right)\right] \nonumber \\
& & - i\pi \, \frac{\sqrt{x-y}}{\sqrt{-y}} \, \left[2 \, \ln 2 + \ln(x-y)\right]
\nonumber \\[0.4cm]
\mbox{\tt R3ex2}(x,y,z) & = & 2 \,\ln\left(1-\frac{x}{z}\right) \, \eta(x,z) + 2
\,\ln\left(1-\frac{y}{z}\right) \, \eta(y,z) \\
& & + 2 \,\Li\left(1-\frac{x}{z}\right) + 2 \,\Li\left(1-\frac{y}{z}\right) + 2
\,(\ln z)^2 \nonumber \\[0.4cm]
\mbox{\tt R3ex3}(x,y,z) & = & 4\,\Spe\left(1-\frac{x}{z}\right) - 4\,\Tri
\left(1-\frac{x}{z}\right) + 4\,\Spe\left(1-\frac{y}{z}\right) - 4\,
\Tri\left(1-\frac{y}{z}\right) \nonumber \\
& & -\,4\,\eta\left(x,\frac{1}{z}\right)\left\{\Li\left(\frac{x}{z}\right) +
\ln\left(1-\frac{x}{z}\right)\ln\left(-\,\frac{x}{z}\right)+\frac{1}{2}
\left[\ln\left(1-\frac{x}{z}\right)\right]^2\right\} \nonumber \\
& & -\,4\,\eta\left(y,\frac{1}{z}\right)\left\{\Li\left(\frac{y}{z}\right) +
\ln\left(1-\frac{y}{z}\right)\ln\left(-\,\frac{y}{z}\right)+\frac{1}{2}
\left[\ln\left(1-\frac{y}{z}\right)\right]^2\right\} \nonumber \\[0.4cm]
\hphantom{\mbox{\tt R3ex3}(x,y,z)} & & -\,4\,\ln z \left[\Li\left(1
-\frac{x}{z}\right)+\Li\left(1-\frac{y}{z}\right)\right] \nonumber \\
& & -\,4\,\ln z \left[\eta\left(x,\frac{1}{z}\right)\ln\left(1-\frac{x}{z}
\right) + \eta\left(y,\frac{1}{z}\right)\ln\left(1-\frac{y}{z}\right)\right]
\nonumber \\
& & + 2\,\Tri\left(\frac{z-y}{z-x}\right) - 2\,\Tri\left(\frac{(z-y)x}{(z-x)y}
\right) + 2\,\Tri\left(\frac{x}{y}\right) \\
& & + \left[\ln\left(\frac{z-y}{z-x}\right)\right]^2\,\eta\left(\frac{y-x}{z},
\frac{z}{z-x}\right) + 2\,\eta\left(y-x,\frac{1}{z}\right) \nonumber \\
& & \qquad\times\left\{\frac{1}{2}\left[\ln\left(
\frac{z-y}{z-x}\right)\right]^2-\ln\left(-\,\frac{z-y}{z-x}\right)
\eta\left(\frac{z-y}{z},\frac{z}{z-x}\right)\right\} \nonumber \\
& & + 2\,(\ln x - \ln y)\left[\Li\left(\frac{(z-y)x}{(z-x)y}\right)-\Li\left(
\frac{x}{y}\right)\right] \nonumber \\
& & - \ln\left(1-\frac{x}{z}\right)\,\left(\ln x - \ln y\right)^2 - 2\,\zeta(3)
- \frac{4}{3} \left(\ln z\right)^3 \nonumber \\
& & + 2 \left[\ln x - \ln y -\frac{1}{2}\ln\left(\frac{(z-y)x}{(z-x)y}\right)
\right]\ln\left(\frac{(z-y)x}{(z-x)y}\right)\eta\left(\frac{y-x}{y},
\frac{z}{z-x}\right) \nonumber
\eea
These functions are related to the corresponding $\mcl R$ functions\footnote{The
function {\tt R3ex3}$(x,y,z)$ assumes that $x$ and $y$ have an imaginary part of
different sign which is always the case.} \cite{Car,Fr3}. They represent the
coefficients of the Taylor expansion in $\eps$ making use of (cf. \cite{Lew})
\beas
\zeta(n) & = & \sum_{k=1}^\infty\,\frac{1}{k^n} \\
\Li(z) & = & - \itg_0^z \frac{\ln(1-s)}{s}\,\,ds \\
\Tri(z) & = & \itg_0^z \frac{\Li(s)}{s}\,\,ds \\
\Spe(z) & = & \frac{1}{2}\,\itg_0^z\frac{\ln^2(1-s)}{s}\,\,ds \\
\eta(a,b) & = & 2\pi i \left[\theta(-\Im a) \theta(-\Im b) \theta(\Im (a b))
- \theta(\Im a) \theta(\Im b) \theta(-\Im (a b))\right]\,.
\eeas
The function \\[0.4cm]
\verb+evalRex(+$\langle${\it expression}$\rangle$\verb+);+ \\[0.4cm]
substitutes the {\tt R{\rm\it n}ex{\rm\it m}} functions in $\langle${\it
expression}$\rangle$ by the corresponding Logarithms and Dilogarithms.

\section{Run time} \label{run}

Due to several optimizations in program code the procedures became much faster
compared to the original version. Run times typical for the new version are
displayed in the following table. We used three different systems: (i) a DEC
Alpha 8400 workstation (ii) a VAX 4000/90 workstation (iii) a PC with i486 chip,
16 MB RAM and 66 MHz frequency. All systems are working with MAPLE~V. Without
using the library we found the following run times:

\bc
\btb{|c|c|c|c|c|c|c|}
\hline
 & \multicolumn{2}{c|}{DEC Alpha 8400} & \multicolumn{2}{c|}{VAX 4000/90} & 
\multicolumn{2}{c|}{PC/DOS} \\
\hline
tensor degree & numerical & algebraic & numerical & algebraic & numerical &
algebraic \\
\hline
\hline
\multicolumn{7}{|c|}{two-point functions} \\
\hline
 0 &  0.02 s & 0.04 s & 0.11 s & 0.22 s & 0.50 s & 0.88 s \\
 1 &  0.04 s & 0.21 s & 0.49 s & 1.53 s & 1.64 s & 2.70 s \\
 2 &  0.07 s & 0.28 s & 0.50 s & 1.97 s & 1.79 s & 3.90 s \\
 3 &  0.08 s & 0.41 s & 0.57 s & 2.99 s & 2.14 s & 8.51 s \\
\hline
\hline
\multicolumn{7}{|c|}{three-point functions} \\
\hline
 0 & 2.54 s & 1.83 s & 19.16 s & 13.82 s &  92.5 s & 25.87 s \\
 1 & 2.57 s & 2.23 s & 20.68 s & 16.66 s & 117.8 s & 26.36 s \\
 2 & 3.01 s & 3.62 s & 20.71 s & 26.31 s & 120.9 s & 38.89 s \\
 3 & 3.78 s & 4.16 s & 28.87 s & 40.30 s & 125.7 s & 117.7 s \\
\hline
\etb \\[0.8cm]
\ec

It should be emphasized that these are the times which are necessary to 
generate the functions once and forever using the {\tt OneLoopLib{\rm\it n}Pt}
routines. If they once are stored in the {\tt LibPath} directory they may be
read in quickly. With assigned {\tt LibPath} we get:

\newpage

\bc
\btb{|c|c|c|c|c|}
\hline
 & \multicolumn{2}{c|}{DEC Alpha 8400}  & \multicolumn{2}{c|}{VAX 4000/90} \\
\hline
tensor degree & numerical & algebraic & numerical & algebraic \\
\hline
\hline
\multicolumn{5}{|c|}{two-point functions} \\
\hline
 0 & 0.34 s & 0.03 s & 2.40 s & 0.09 s \\
 1 & 0.39 s & 0.05 s & 2.95 s & 0.15 s \\
 2 & 0.43 s & 0.06 s & 3.22 s & 0.19 s \\
 3 & 0.47 s & 0.06 s & 3.43 s & 0.21 s \\
\hline
\hline
\multicolumn{5}{|c|}{three-point functions} \\
\hline
 0 & 1.30 s & 0.09 s &  9.27 s & 0.45 s \\
 1 & 1.32 s & 0.10 s &  9.97 s & 0.48 s \\
 2 & 1.36 s & 0.12 s & 10.30 s & 0.56 s \\
 3 & 1.88 s & 0.45 s & 12.08 s & 0.65 s \\
\hline
\etb \\[0.8cm]
\ec

\section*{Appendix: Installation and program call}

The distribution of the {\it oneloop} package is accessible at \\[0.4cm]
\verb+http://wwwthep.physik.uni-mainz.de/~xloops/+ \\[0.4cm]
To install the package, just copy all files with the extension {\tt .ma} in one
directory. After having started the MAPLE session, you can read the {\it
oneloop} package -- just type the following lines: \\[0.4cm]
\verb+LoopPath:=+{$\langle$\it path$\rangle$}\verb+;+ \\
\verb+read`+{$\langle$\it path$\rangle$}\verb+oneloop.ma`;+ \\[0.4cm]
{$\langle$\it path$\rangle$} describes the directory where the {\tt .ma} files
are located. If these files are in the current directory then you can skip the
assignment of \verb+LoopPath+ and only type \\[0.4cm]
\verb+read`oneloop.ma`;+ 

\section*{Acknowledgements}

We would like to thank A. Frink for his check of ${\mcl O}(\eps)$ contributions.


\begin{thebibliography}{10}

\bibitem{Ma1}
{B. W. Char, K. O. Geddes, G. H. Gonnet, B. L. Leong, M. B. Monagan, S. M.
  Watt}.
\newblock {\titlestyle {{\sf Maple V} Language Reference Manual}}.
\newblock Springer (1991)

\bibitem{Ma2}
{B. W. Char, K. O. Geddes, G. H. Gonnet, B. L. Leong, M. B. Monagan, S. M.
  Watt}.
\newblock {\titlestyle {{\sf Maple V} Library Reference Manual}}.
\newblock Springer (1991)

\bibitem{Fr4}
{L. Br\"ucher, J. Franzkowski, D. Kreimer}.
\newblock
\newblock {\titlestyle Comp. Phys. Comm.} {\volstyle 85} (1995) 153

\bibitem{Col}
{J. C. Collins}.
\newblock {\titlestyle {Renormalization}}.
\newblock Cambridge University Press (1984)

\bibitem{Fr13}
{L. Br\"ucher, J. Franzkowski, A. Frink, D. Kreimer}.
\newblock
\newblock {\titlestyle hep-ph/{\bf 3611378}, Univ. Mainz Preprint} {\volstyle
  MZ-TH/96-39} (1996)

\bibitem{Nie}
{U. Nierste, D. M\"uller, M. B\"ohm}.
\newblock
\newblock {\titlestyle Z. Phys.} {\volstyle C57} (1993) 605

\bibitem{Fr9}
{L. Br\"ucher, J. Franzkowski}.
\newblock {\titlestyle {Two-loop Relevant Parts of One-loop Functions}}.
\newblock \prepname

\bibitem{Gas}
{R. Gastmans, J. Verwaest, R. Meuldermans}.
\newblock
\newblock {\titlestyle Nucl. Phys.} {\volstyle B105} (1976) 454

\bibitem{Pas}
{G. Passarino, M. Veltman}.
\newblock
\newblock {\titlestyle Nucl. Phys.} {\volstyle B160} (1979) 151

\bibitem{Car}
{B. C. Carlson}.
\newblock {\titlestyle {Special Functions of Applied Mathematics}}.
\newblock Academic Press (1977)

\bibitem{Fr3}
{L. Br\"ucher, J. Franzkowski, D. Kreimer}.
\newblock
\newblock {\titlestyle Mod. Phys. Lett.} {\volstyle A9} (1994) 2335

\bibitem{Lew}
{L. Lewin}.
\newblock {\titlestyle {Polylogarithms and Associated Functions}}.
\newblock North Holland (1981)

\end{thebibliography}

\end{document}